\documentclass[conference]{IEEEtran}
\IEEEoverridecommandlockouts
\usepackage{cite}
\usepackage{amsmath,amssymb,amsfonts}
\usepackage{algorithmic}
\usepackage{graphicx}
\usepackage{textcomp}
\usepackage{xcolor}
\usepackage[nolist]{acronym}
\def\BibTeX{{\rm B\kern-.05em{\sc i\kern-.025em b}\kern-.08em
    T\kern-.1667em\lower.7ex\hbox{E}\kern-.125emX}}
\begin{document}

\sloppy
\begin{acronym}

\acro{2D}{Two Dimensions}%
\acro{2G}{Second Generation}%
\acro{3D}{Three Dimensions}%
\acro{3G}{Third Generation}%
\acro{3GPP}{Third Generation Partnership Project}%
\acro{3GPP2}{Third Generation Partnership Project 2}%
\acro{4G}{Fourth Generation}%
\acro{5G}{Fifth Generation}%

\acro{AI}{Artificial Intelligence}%
\acro{AoA}{Angle of Arrival}%
\acro{AoD}{Angle of Departure}%
\acro{AR}{Augmented Reality}%
\acro{AP}{Access Point}
\acro{AE}{Antenna Element}
\acro{AC}{Anechoic Chamber}
\acro{AUT}{Antenna Under Test}

\acro{BER}{Bit Error Rate}%
\acro{BPSK}{Binary Phase-Shift Keying}%
\acro{BRDF}{ Bidirectional Reflectance Distribution Function}%
\acro{BS}{Base Station}%

\acro{CA}{Carrier Aggregation}%
\acro{CDF}{Cumulative Distribution Function}%
\acro{CDM}{Code Division Multiplexing}%
\acro{CDMA}{Code Division Multiple Access}%
\acro{CPU} {Central Processing Unit}
\acro{CUDA}{Compute Unified Device Architecture}
\acro{CDF}{Cumulative Distribution Function}
\acro{CI}{Confidence Interval}
 
\acro{D2D}{Device-to-Device}%
\acro{DL}{Down Link}%
\acro{DS}{Delay Spread}%
\acro{DAS}{Distributed Antenna System}
\acro{DKED}{double knife-edge diffraction}
\acro{DUT}{Device Under Test}


\acro{EDGE}{Enhanced Data rates for GSM Evolution}%
\acro{EIRP}{Equivalent Isotropic Radiated Power}%
\acro{eMBB}{Enhanced Mobile Broadband}%
\acro{eNodeB}{evolved Node B}%
\acro{ETSI}{European Telecommunications Standards Institute}%
\acro{ER}{Effective Roughness}%
\acro{E-UTRA}{Evolved UMTS Terrestrial Radio Access}%
\acro{E-UTRAN}{Evolved UMTS Terrestrial Radio Access Network}%
\acro{EF}{Electric Field}

\acro{FDD}{Frequency Division Duplexing}%
\acro{FDM}{Frequency Division Multiplexing}%
\acro{FDMA}{Frequency Division Multiple Access}%
\acro{FoM}{Figures of Merit}

\acro{GI}{Global Illumination} %
\acro{GIS}{Geographic Information System}%
\acro{GO}{Geometrical Optics} %
\acro{GPU}{Graphics Processing Unit}%
\acro{GPGPU}{General Purpose Graphics Processing Unit}%
\acro{GPRS}{General Packet Radio Service}%
\acro{GSM}{Global System for Mobile Communication}%

\acro{H2D}{Human-to-Device}%
\acro{H2H}{Human-to-Human}%
\acro{HDRP}{High Definition Render Pipeline}
\acro{HSDPA}{High Speed Downlink Packet Access}
\acro{HSPA}{High Speed Packet Access}%
\acro{HSPA+}{High Speed Packet Access Evolution}%
\acro{HSUPA}{High Speed Uplink Packet Access}
\acro{HPBW}{Half-Power Beamwidth}

\acro{IEEE}{Institute of Electrical and Electronic Engineers}%
\acro{InH}{Indoor Hotspot} %
\acro{IMT} {International Mobile Telecommunications}%
\acro{IMT-2000}{\ac{IMT} 2000}%
\acro{IMT-2020}{\ac{IMT} 2020}%
\acro{IMT-Advanced}{\ac{IMT} Advanced}%
\acro{IoT}{Internet of Things}%
\acro{IP}{Internet Protocol}%
\acro{ITU}{International Telecommunications Union}%
\acro{ITU-R}{\ac{ITU} Radiocommunications Sector}%
\acro{IS-95}{Interim Standard 95}%
\acro{IES}{Inter-Element Spacing}


\acro{KPI}{Key Performance Indicator}%
\acro{K-S}{Kolmogorov-Smirnov}

\acro{LB} {Light Bounce}
\acro{LIM}{Light Intensity Model}%
\acro{LoS}{line of sight}%
\acro{LTE}{Long Term Evolution}%
\acro{LTE-Advanced}{\ac{LTE} Advanced}%
\acro{LSCP}{Lean System Control Plane}%
\acro{LSI} {Light Source Intensity}

\acro{M2M}{Machine-to-Machine}%
\acro{MatSIM}{Multi Agent Transport Simulation}
\acro{METIS}{Mobile and wireless communications Enablers for Twenty-twenty Information Society}%
\acro{METIS-II}{Mobile and wireless communications Enablers for Twenty-twenty Information Society II}%
\acro{MIMO}{Mul\-ti\-ple-In\-put Mul\-ti\-ple-Out\-put}
\acro{mMIMO}{massive MIMO}%
\acro{mMTC}{massive Machine Type Communications}%
\acro{mmW}{millimeter-wave}%
\acro{MU-MIMO}{Multi-User MIMO}
\acro{MMF}{Max-Min Fairness}
\acro{MKED}{Multiple Knife-Edge Diffraction}
\acro{MF}{Matched Filter}
\acro{mmWave}{Millimeter Wave}

\acro{NFV}{Network Functions Virtualization}%
\acro{NLoS}{non line of sight}%
\acro{NR}{New Radio}%
\acro{NRT}{Non Real Time}%
\acro{NYU}{New York University}%

\acro{O2I}{Outdoor to Indoor}%
\acro{O2O}{Outdoor to Outdoor}%
\acro{OFDM}{Orthogonal Frequency Division Multiplexing}%
\acro{OFDMA}{Or\-tho\-go\-nal Fre\-quen\-cy Di\-vi\-sion Mul\-ti\-ple Access}
\acro{OtoI}{Outdoor to Indoor}%
\acro{OTA}{Over-The-Air}

\acro{PDF}{Probability Distribution Function}
\acro{PDP}{Power Delay Profile}
\acro{PHY}{Physical}%
\acro{PLE}{Path Loss Exponent}

\acro{QAM}{Quadrature Amplitude Modulation}%
\acro{QoS}{Quality of Service}%

\acro{RCSP}{Receive Signal Code Power}
\acro{RAN}{Radio Access Network}%
\acro{RAT}{Radio Access Technology}%

\acro{RAN}{Radio Access Network}%
\acro{RMa}{Rural Macro-cell}%
\acro{RMSE} {Root Mean Square Error}
\acro{RSCP}{Receive Signal Code Power}%
\acro{RT}{Ray Tracing}
\acro{RX}{receiver}
\acro{RMS}{Root Mean Square}
\acro{Random-LOS}{Random Line-Of-Sight}
\acro{RF}{Radio Frequency}
\acro{RC}{Reverberation Chamber}
\acro{RIMP}{Rich Isotropic Multipath}

\acro{SB} {Shadow Bias}
\acro{SC}{small cell}
\acro{SDN}{Software-Defined Networking}%
\acro{SGE}{Serious Game Engineering}%
\acro{SF}{Shadow Fading}%
\acro{SIMO}{Single Input Multiple Output}%
\acro{SINR}{Signal to Interference plus Noise Ratio}
\acro{SISO}{Single Input Single Output}%
\acro{SMa}{Suburban Macro-cell}%
\acro{SNR}{Signal to Noise Ratio}
\acro{SU}{Single User}%
\acro{SUMO}{Simulation of Urban Mobility}
\acro{SS} {Shadow Strength}


\acro{TDD}{Time Division Duplexing}%
\acro{TDM}{Time Division Multiplexing}%
\acro{TD-CDMA}{Time Division Code Division Multiple Access}%
\acro{TDMA}{Time Division Multiple Access}%
\acro{TX}{transmitter}
\acro{TZ}{Test Zone}
\acro{TRP}{Total Radiated Power}


\acro{UAV}{Unmanned Aerial Vehicle}%
\acro{UE}{User Equipment}%
\acro{UI}{User Interface}
\acro{UHD}{Ultra High Definition}
\acro{UL}{Uplink}%
\acro{UMa}{Urban Macro-cell}%
\acro{UMi}{Urban Micro-cell}%
\acro{uMTC}{ultra-reliable Machine Type Communications}%
\acro{UMTS}{Universal Mobile Telecommunications System}%
\acro{UPM}{Unity Package Manager}
\acro{UTD}{Uniform Theory of Diffraction} %
\acro{UTRA}{{UMTS} Terrestrial Radio Access}%
\acro{UTRAN}{{UMTS} Terrestrial Radio Access Network}%
\acro{URLLC}{Ultra-Reliable and Low Latency Communications}%

\acro{V2V}{Vehicle-to-Vehicle}%
\acro{V2X}{Vehicle-to-Everything}%
\acro{VP}{Visualization Platform}%
\acro{VR}{Virtual Reality}%
\acro{VNA}{Vector Network Analyzer}
\acro{VIL}{Vehicle-in-the-loop}

\acro{WCDMA}{Wideband Code Division Multiple Access}%
\acro{WINNER}{Wireless World Initiative New Radio}%
\acro{WINNER+}{Wireless World Initiative New Radio +}%
\acro{WiMAX}{Worldwide Interoperability for Microwave Access}%
\acro{WRC}{World Radiocommunication Conference}%

\acro{xMBB}{extreme Mobile Broadband}%

\acro{ZF}{Zero Forcing}

\end{acronym}

\title{Total Radiated Power Measurements of a mmWave Phased Array in a Reverberation Chamber\\

\thanks{The work of Alejandro Antón is supported by the European Union’s Horizon 2020 Marie Skłodowska-Curie grant agreement No. 955629. Andrés Alayón Glazunov also kindly acknowledges funding from the ELLIIT strategic research environment (https://elliit.se/).}
}

\author{\IEEEauthorblockN{Alejandro Antón Ruiz}
\IEEEauthorblockA{\textit{Department of Electrical Engineering} \\
\textit{University of Twente}\\
Enschede, Netherlands \\
a.antonruiz@utwente.nl}
\and
\IEEEauthorblockN{Samar Hosseinzadegan,\\ John Kvarnstrand, Klas Arvidsson}
\IEEEauthorblockA{\textit{Bluetest AB}\\
Gothenburg, Sweden \\
name.familyname@bluetest.se}
\and
\IEEEauthorblockN{Andrés Alayón Glazunov}
\IEEEauthorblockA{\textit{University of Twente} \\
\textit{Department of Science and Technology} \\
\textit{Linköping University}\\
Norrköping Campus, Sweden \\
andres.alayon.glazunov@liu.se}
}
\maketitle

\begin{abstract}
This paper explores the use of reverberation chambers for TRP measurements of beamformed radiation by phased arrays at mmWave frequencies. First, the received power was verified by the one-sample K-S GoF test to follow the exponential probability distribution. Different numbers of samples and stirrers’ positions were considered. Second, we showed that the effective number of independent samples is different depending on the number of samples and stirrers’ positions. Third, the beamforming TRP estimates are presented for all beams, analyzing the statistical significance of the observed differences with a selection of samplings.
\end{abstract}


\section{Introduction}
Phased arrays are a 5G enabling technology exploiting the \ac{mmWave} spectrum, overcoming the propagation loss. \ac{OTA} performance evaluation is a key aspect in developing, producing, and deploying antennas. The radiated performance of a wireless device is fundamentally described by its \ac{TRP}. For phased arrays \ac{TRP} is important from the energy efficiency point of view and also due to regulatory and interference restrictions. The quality of the \ac{AUT} can be assured, e.g., by evaluating \ac{TRP} performance consistency among beams or comparing against a golden device. It is widely known that an \ac{RC} is much more efficient in evaluating \ac{TRP} as compared to anechoic chamber measurements. Works presenting \ac{mmWave} phased arrays measurements in \acp{RC} are not new \cite{5GTILERC,6GAERC}. However, to the best knowledge of the authors, there are no works of \ac{TRP} measurements of active phased arrays at \ac{mmWave} in \acp{RC}. Thus, this work aims at filling this gap.

\section{Measurement setup and procedure}
\subsection{Phased array antenna}
An evaluation kit (EVK02001), with \ac{RF} module (BFM02003) from Sivers Semiconductors AB and its own continuous wave source restrictions, is used in this study \cite{Sivers}. It operates between $24-29.5$~GHz. The BFM02003 has two identical modules, each with $2$ rows of $8$ linearly polarized patch \acp{AE}. The RX and the TX have 16 \acp{AE} each. Only the TX module is used here and evaluated at $28$~GHz. The azimuth scanning covers from $-45^{\circ}$ to $+45^{\circ}$ with $4.5^{\circ}$ step size. Thus, it produces 21 different beams, denoted as beam 1 starting from $-45^{\circ}$ and so on.

\subsection{Reverberation chamber and instruments}

The \ac{OTA} \ac{TRP} experiments were performed with an RTS65 \ac{RC} from Bluetest AB. RTS65 is capable of \ac{mmWave} measurements up to $43.5$~GHz and is equipped with two linearly polarized measurement antennas. Here, only one of them was used. Note that the \ac{RC} achieves polarization balance. \ac{TRP} calibration requires measuring the chamber loss. This was done using a reference horn antenna and a \ac{VNA}, keeping inside the \ac{RC} the phased array and its fixture. Then, the \ac{TRP} measurement of the \ac{AUT} was performed with chamber loss compensation, using a signal and spectrum analyzer. This was repeated for the 21 beams collecting 600 samples per beam. Each sample corresponds to different positions of two different mechanical stirrers, i.e., a turntable moving in the outer loop and with a $14.4^{\circ}$ step size, and vertical and horizontal metal paddles, moving in the inner loop. Beam switching was done in the outermost loop. Specifically, the first 25 samples differ in paddle positions and share the same turntable position. The next 25 samples share a new position of the turntable and so on. After 600 samples are taken, beam switching occurs.

\section{Results}
Various numbers of samples were considered: 100, 150, 200, 300, and 600. The smaller sample subsampling sets are taken from the same initial 600 samples. Two subsampling approaches were considered: (i) contiguous (C) (e.g., the first 300 samples are kept), and (ii) decimated (D) (e.g., only every 2$^{\textrm{nd}}$ sample is kept starting from the first one). Note that subsampling (i) does not cover the whole circle ($0^{\circ}$ to $331.2^{\circ}$) of the turntable rotation while subsampling (ii) does. The gathered samples are [1 2…600] (600), [1 2…300] (300C), [1 3…599] (300D), [1 2…200] (200C), [1 4…598] (200D), [1 2…150] (150C), [1 5…597] (150D), [1 2…100] (100C), [1 7…595] (100D), where C and D denote subsampling-(i) and –(ii), respectively.

\subsection{Goodness of fit test}

The \ac{PDF} of the received signal power measured in a well-stirred \ac{RC} must follow the exponential distribution. The one-sample \ac{K-S} test is applied at the $5\%$ significance level. For all the considered samplings, the test could not reject the hypothesis that the samples come from an exponential distribution. Thus, even if the turntable rotation is partial, e.g., the 100C case covering $43.2^\circ$ of the turntable rotation, the \ac{PDF} is most likely exponential.

\subsection{Independent samples}

\begin{figure}
    \centering
        \includegraphics[width=0.76\columnwidth]{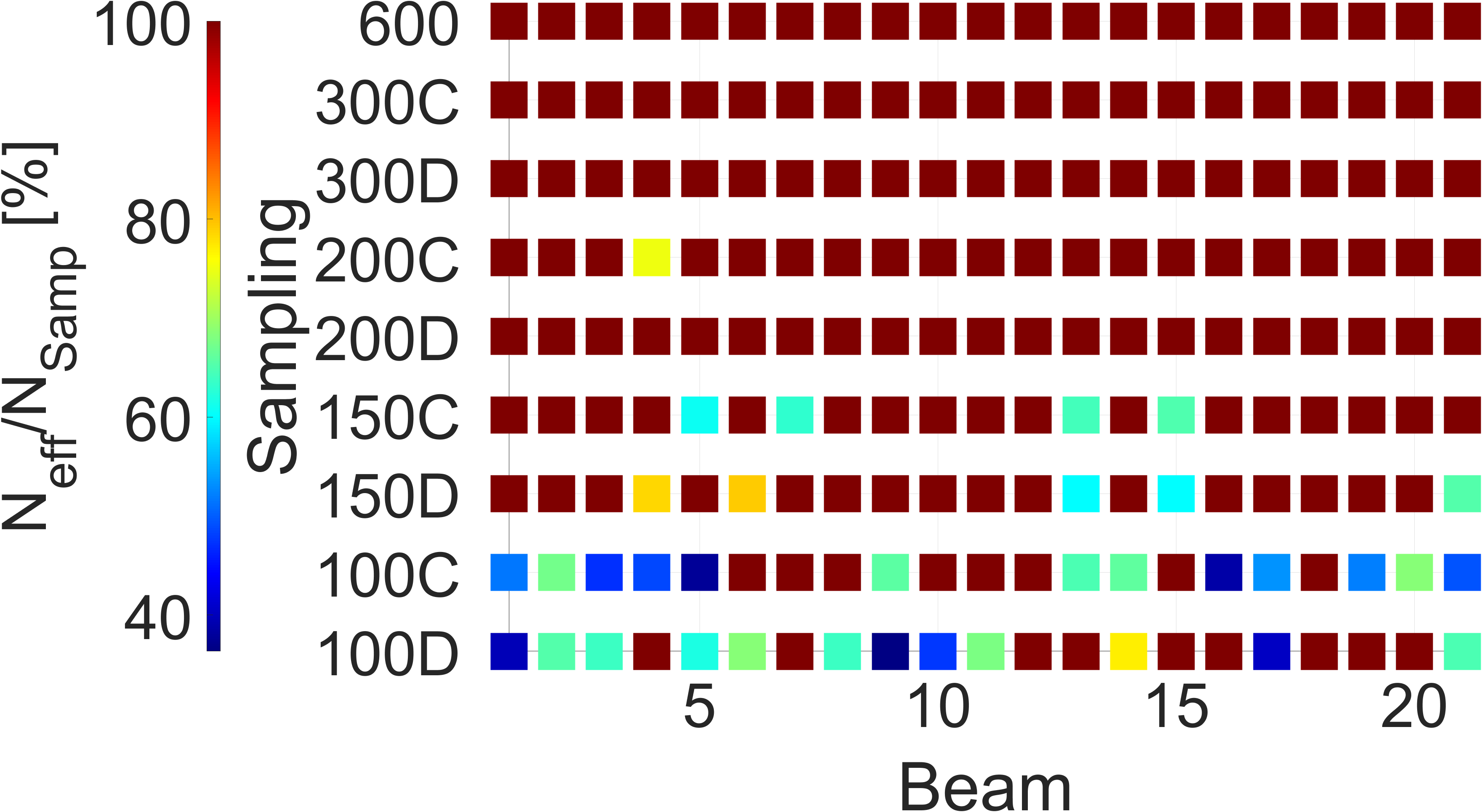}
    \caption{Percentage of $N_{eff}/N_{Samp}$ for all samplings and all beams.}
    \label{F1}
\end{figure}

IEC and 3GPP standards define how to measure the number of effective samples \cite{IEC61000421,3GPPRCStandard}. The definitions are only valid for one tuner and equidistant tuner positions though. In \cite{CorrMatrix}, a circular-shift correlation matrix approach, suitable for more than one tuner compatible with \cite{IEC61000421,3GPPRCStandard}, is proposed. This approach is used here to produce results shown in Fig.~\ref{F1}, which indicate the percentage of effective samples $N_{eff}$, i.e., independent samples, relative to the considered samples $N_{Samp}$. As can be seen, reducing $N_{Samp}$ implies a general reduction of $N_{eff}/N_{Samp}$. This is because the threshold established in \cite{IEC61000421,3GPPRCStandard}, also used in \cite{CorrMatrix}, valid for $N_{Samp}\geq100$ at $95\%$ confidence level, becomes stricter with decreasing $N_{Samp}$. The relative accuracy of the \ac{TRP} estimate, denoted as $\widehat{TRP}$, has a standard deviation given by $\sigma=1/\sqrt{N_{eff}}$ \cite{LTEBOOK}. Thus, the $\widehat{TRP}$ from sets of samples with $N_{eff}/N_{Samp}$ lower than 100\%, will have a larger uncertainty at a given $N_{Samp}$. For this particular case, if one wants to assure a given level of $\sigma$ governed by $N_{Samp}$ ($N_{eff}=N_{Samp}$), then $N_{Samp}$ will need to be at least 300 (or 200 using the 200D sampling).

\subsection{TRP statistics}
\begin{figure}
    \centering
        \includegraphics[width=0.76\columnwidth]{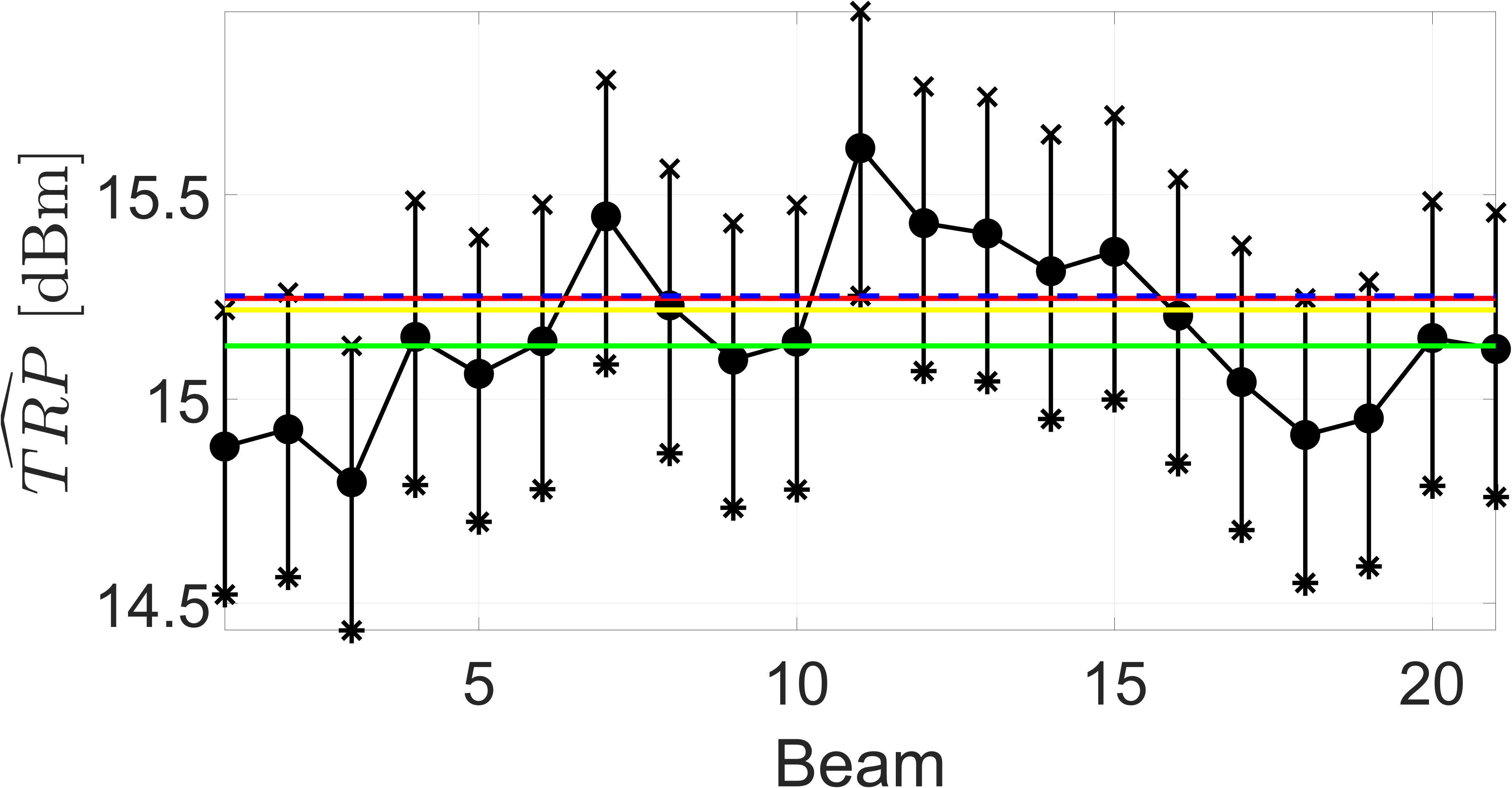}
    \caption{In black, $\widehat{TRP}$ ($\bullet$) with $95\%$ \ac{CI} upper ($\times$) and lower ($\ast$) bounds for 600 sampling and all beams. $95\%$ lower bound of beam 11 in blue. $95\%$ upper bounds of beams 1 (yellow), 3 (green), and 18 (red).}
    \label{F2}
\end{figure}

Fig.~\ref{F2} shows the $\widehat{TRP}$ and the $95\%$ \ac{CI} for the 600 sampling, which has $N_{eff}=N_{Samp}$ for all beams. To affirm that there is a significant statistical difference (with $95\%$ confidence) between values of $\widehat{TRP}$ from different beams, no overlap between their respective \acp{CI} can exist, i.e., that the \ac{CI} upper bound of a given $\widehat{TRP}$ is larger than the \ac{CI} lower bound of other $\widehat{TRP}$. For almost all cases, there is such overlap, meaning that there is no statistically significant difference at $95\%$ confidence level between most beams. Specifically, for the 600 sampling, only the $\widehat{TRP}$ of beam 11 is larger than that of beams 1, 3, and 18 (see Fig.~\ref{F2}). In case we use the 300D sampling only the $\widehat{TRP}$ of beam 3 is lower than in beams 11, 12, and 13. Finally, for the 300C sampling, the $\widehat{TRP}$ of beam 11 is larger than in beams 2, 17, 18, and 19, and $\widehat{TRP}$ of beam 7 is larger than in beam 18. Thus, the different samplings have an influence in the $\widehat{TRP}$ statistically significant differences, but, overall, most $\widehat{TRP}$ values of different beams cannot be considered different from each other. This indicates a good performance of the \ac{AUT}, since a phased array should have the same \ac{TRP} for all beams.

\section{Conclusions}

This paper has presented an analysis of per beam \ac{TRP} measurements of an active phased array in an \ac{RC}. The data can be assumed to be measured in \ac{RIMP}. It has also been shown that $N_{eff}/N_{Samp}$ depends on the sampling procedure and no statistically significant differences of $\widehat{TRP}$ have been found for most beams.

\bibliographystyle{IEEEtran}

\bibliography{References}

\end{document}